\documentclass[fleqn,10pt]{olplainarticle}

\title{Deep or Not Deep: Supervised Learning Approaches to Modeling the Pedestal Density}

\author[1]{Adam Kit}
\author[1, 2]{Aaro Järvinen}
\author[3]{Lorenzo Frassinetti}
\author[4]{Sven Wiesen}
\author[5]{JET Contributors}
\affil[1]{University of Helsinki, FI-00014 Helsinki, Finland; adamkit@helsinki.fi}
\affil[2]{VTT Technical Research Centre of Finland, FI-02044 VTT, Finland}
\affil[3]{KTH Royal Institute of Technology, SE-11428 Stockholm, Sweden}
\affil[4]{Forschungszentrum Jülich GmbH, Institut für Energie- und Klimaforschung– Plasmaphysik, DE-52425 Jülich, Germany}
\affil[5]{See the author list of ‘Overview of JET results for optimising ITER operation’ by J. Mailloux et al. to be
published in Nuclear Fusion Special issue: Overview and Summary Papers from the 28th Fusion Energy
Conference (Nice, France, 10-15 May 2021)}

\begin{abstract}
Pedestal is the key to conventional high performance plasma scenarios in tokamaks. However, high fidelity simulations of pedestal plasmas are extremely challenging due to the multiple physical processes and scales that are encompassed by tokamak pedestals. The leading paradigm for predicting the pedestal top pressure is encompassed by EPED-like models \cite{Snyder_2009, Snyder_2011, Snyder_2012, Saarelma_2019, Dunne_2017}. However, EPED does not predict the pedestal top density, $n_\text{e,ped}$, but requires it as an input. EUROPED \cite{Saarelma_2019} employs simplified models, such as log-linear regression, to constrain $n_\text{e,ped}$ with tokamak machine control parameters in an EPED-like model. However, these simplified models for $n_\text{e,ped}$ often show disagreements with experimental observations and do not use all of the available numerical and categorical machine control information. In this work it is observed that using the same input parameters, decision tree ensembles and deep learning models improve the predictive quality of $n_\text{e,ped}$ by about 23\% relative to that obtained with log-linear scaling laws, measured by root mean square error. Including all of the available tokamak machine control parameters, both numerical and categorical, leads to further improvement of about 13\%. Finally, predictive quality was tested when including global normalized plasma pressure and effective charge state as inputs, as these parameters are known to impact pedestals. Surprisingly, these parameters lead to only a few percent further improvement of the predictive quality.

\end{abstract}

\begin{document}

\flushbottom
\maketitle
\thispagestyle{empty}

\section{Introduction}

One of the primary challenges faced by future fusion reactors is to integrate high performance fusion plasmas with tolerable conditions at the reactor components \citep{Ikeda_2007, Zohm_2017}. The conventional approach toward a high performance fusion core in tokamak plasmas is to operate the plasma in a high confinement mode (H-mode), such that a transport barrier and pedestal are formed at the edge of the plasma \cite{Zohm_1996}. The tension of core-edge integration balances at the pedestal, with mutually competing requirements of high pedestal pressure for fusion performance and mitigation of plasma heat fluxes in the scrape-off layer to avoid component overheating. Therefore, predictive capability for the pedestal region of the tokamak plasma is essential for predicting performance and core-edge integration in future fusion reactors. 

First principles simulations for pedestal plasmas are extremely challenging, and the ongoing effort of the scientific community has led to models such as EPED \cite{Snyder_2009, Snyder_2011, Snyder_2012}, EUROPED \cite{Saarelma_2017, Saarelma_2019}, and iPED \cite{Dunne_2017}  that predict the pedestal height and width based on certain simplifying assumptions. The EPED-like models encompass the leading paradigm for predicting pedestal top pressures assuming that the pedestal gradient is turbulence limited by the kinetic ballooning mode and the pedestal top pressure is magnetohydrodynamic  stability limited by the peeling-ballooning mode \cite{Snyder_2009, Snyder_2011, Snyder_2012}. However, EPED cannot be considered a fully predictive model as, in addition to machine control parameters, EPED applies information about the confined plasma state, such as total normalized plasma plasma pressure relative to the magnetic pressure, $\beta_N$, and pedestal top density, $n_\text{e,ped}$, as inputs. EUROPED \cite{Saarelma_2019} can work exactly as EPED or can additionally employ simple core transport models to constrain $\beta_N$ and models for $n_\text{e,ped}$. With these additional models, EUROPED is able to predict pedestal plasmas with an EPED-like model using only engineering control parameters. However, both models for $n_\text{e,ped}$ that are employed by EUROPED, (i) the log-linear regression given by Urano \cite{Urano} based on experimental data and (ii) the neutral penetration model, show disagreements with experimental observations for the JET database, especially for $n_\text{e,ped}$ exceeding $8\times10^{19}$ m$^{-3}$ \cite{Saarelma_2019}.

Supervised machine learning (ML) offers potential for regression beyond the standard log-linear approach to predict $n_\text{e,ped}$ used by EUROPED. In this work, various ML models are applied for predicting $n_\text{e,ped}$ using the JET pedestal database described in \citep{Frassinetti_2020}, and the performance of these models is compared to the log-linear scaling laws for $n_\text{e,ped}$ produced by Urano \citep{Urano} and Frassinetti \citep{Frassinetti_2020}. A related study by Gillgren et al. \citep{Gillgren_2021}, investigated the  application of shallow artificial neural networks (ANNs) with promising results for pedestal electron density and temperature predictions to provide boundary conditions for core transport simulations. 

With this work we aim to answer the following questions:
\begin{enumerate}
    \item \textit{\textbf{What is the hierarchy of input parameters in terms of providing information to improve the pedestal density prediction quality?}}
    The current state-of-the-art log-linear scaling laws from Urano \citep{Urano} and Frassinetti \citep{Frassinetti_2020} use reduced subsets of all available machine control parameters as inputs. These subsets were chosen as the leading order input parameters using domain knowledge. Machine learning algorithms can utilize all possible input data, even if there are strong cross-correlations among the inputs. Therefore, it is important to compare the predictive performance of the ML algorithms when the input space is expanded to include (i) all the available numerical machine control parameters, (ii) categorical machine control parameters, such as divertor configuration, and (iii) global plasma information through dimensionless normalized plasma pressure, $\beta_\text{N}$, and effective charge, $Z_\text{eff}$, that are known to impact the performance of the pedestal.

    
    \item 
    \textit{\textbf{Can advanced machine learning methods like tree-based ensembles and deep learning improve prediction quality for pedestal density predictions?}}
    ML can encompass a wide variety of approaches, ranging from traditional linear regression to more complex deep learning networks or decision tree ensemble methods. It is not always \textit{a priori} obvious which of these approaches is best suited for the given application. Additionally, with increased complexity, the time for training and application (inference) increases. Even though it would intuitively seem that more complex models would increase the risk of overfitting by having the capacity to memorize samples, over-parameterized deep learning networks tend to still maintain generalization capabilities \citep{Generalization}. This work aims to document the advantages and disadvantages of these approaches when developing a regression based predictor for $n_\text{e,ped}$.     
\end{enumerate}

\section{JET Pedestal Database}

\newlength{\oldintextsep}
\setlength{\oldintextsep}{\intextsep}
\setlength\intextsep{0pt}

The data used in this work is obtained from the EUROfusion JET pedestal database \cite{Frassinetti_2020}. A full analysis of the database can be found in \cite{Frassinetti_2020}. The data is in tabular form with rows (entries) of shots and time windows that have columns (features) corresponding to time averaged machine control parameters and measurements of various plasma parameters. From 2625 unique JET shots, the dataset consists of 3557 entries.

The subset of the JET pedestal database used in this analysis includes techniques to control edge localized modes, such as kicks \citep{Kicks_2015} or resonant magnetic perturbations (RMPs), \citep{RMP_2006}, as well as plasmas with impurity seeding and pellets \cite{Pellets_1997, Giroud_2015}. We can feed this information to machine learning models in a numerically encoded format, such that shots with kicks are given a column value 1 and shots without a column value 0. This approach extends to an arbitrary number of categories as the list can be extended to as many numbers as are needed to describe the category.
While the included dataset contains over 1500 shots with the JET ITER-like tungsten beryllium wall (JET-ILW) \cite{MatthewsPSC2011}, there are also 422 shots with the previous carbon wall (JET-C). While a dedicated wall material flag is not given to the algorithms, the information about JET-C is mostly coded in the categorical variable containing the charge number of the impurity species in the plasma, $Z_\text{imp}^\text{atomic}$. However, there are 21 shots with JET-ILW and $Z_\text{imp}^\text{atomic} = 6$, due to seeding of methane. Future studies will investigate the impact of including a dedicated wall material flag. Furthermore, a dedicated impurity injection rate value is not given in the input set, which is expected to reduce predictive performance as the database contains plasmas at varying impurity injection rates. The role of the impurity injection rate will be addressed in future studies. Even though impurity injection rate is not included explicitly, this information is partially coded into $Z_\text{eff}$ and, therefore, accessible to model experiments containing $Z_\text{eff}$  information as input. The major radius, $R$, is notably missing from the list of main engineering parameters found in Table \ref{tab:parameter_domains}. For single machine analysis, $R$ does not vary in a meaningful way, and through the Shafranov shift of the magnetic configuration \citep{Shafranov_1974}, encodes data about the pedestal pressure. 
 \begin{table}[htb]   
 \caption{Relevant parameter domains of JET pedestal database used in this analysis. Individual parameters are detailed in \cite{Frassinetti_2020}. Features with $\dagger, \ddagger$ correspond to those used in the Urano and Frassinetti subsets of input space respectively. While for triangularity, $\delta$, Urano and Frassinetti used the average value, here separate values are used for the upper and lower triangularities. The divertor configuration categorical variable represents the specific divertor plate of the strike points. A plasma with a 'V/H' divertor configuration has an inner strike point on the vertical target and outer strike point on the horizontal target, and is detailed in Figure 5 of \cite{Frassinetti_2020}. $P_{NBI}$ refers to the applied heating power from neutral beam injection.}
    \centering
    \begin{tabular}{| c | c |}
    \hline
        Feature (Units) & Domain [Min, Max] \\
        \hline
        \multicolumn{2}{c}{Main Engineering} \\
        \hline
        $I_P$ (MA) $\dagger, \ddagger$& [0.813, 4.48] \\
        $B_T$  (T) $\dagger$ & [0.9616, 3.685] \\
        $a$ (m) & [0.8278, 0.9749] \\ 
        $P_\text{NBI}$ (MW) $\dagger$& [0, 32.1367] \\ 
        $P_\text{ICRH}$ (MW)& [0, 7.963] \\
        $P_\text{TOT}$ (MW) $\ddagger$& [3.4021, 38.2205] \\
        $V_P$ (m$^3$) & [58.301, 82.1878] \\ 
        $q_{95}$ (-) & [2.425, 6.095] \\
        $\delta_\text{lower}$ (-) $\dagger, \ddagger$& [0.2318, 0.4915] \\
        $\delta_\text{upper}$ (-) $\dagger, \ddagger$& [0.0438,0.5896 ] \\
        $\kappa$ (-) & [1.587, 1.819] \\
        $\Gamma$ ($10^{22}$ e/s) $\dagger, \ddagger$ & [0, 15.502] \\
        \hline
        \multicolumn{2}{c}{Categorical} \\
        \hline
        Divertor &  [C/C,V/H,V/C,V/V,C/V,C/H] \\ 
        $Z_\text{imp}^\text{atomic}$ & [0, 2, 6, 7, 8, 10, 18, 36] \\ 
        Kicks & [0, 1] \\ 
        RMPs & [0, 1] \\ 
        Pellets & [0, 1] \\ 
        \hline
        \multicolumn{2}{c}{Global} \\
        \hline
        $\beta_\text{N}^\text{MHD}$ (-) & [0.574, 3.625] \\
        $Z_\text{eff}$ (-) & [1, 3.746] \\
        \hline
        \multicolumn{2}{c}{Target} \\
        \hline
        $n_\text{e,ped}$ ($10^{19}$ m$^{-3}$) & [1.4192, 11.737] \\
        \hline
    \end{tabular}
   
\label{tab:parameter_domains}
\end{table}
\setlength\intextsep{\oldintextsep}

\section{Machine Learning for Tabular Datasets}
Decision tree ensembles and deep learning methods can capture many nuances of tabular datasets. A decision tree ensembles can exploit multiple decision trees fit on unique subsets of the datasets while deep learning models make use of specialized layers designed for tabular data. A brief description of each model used and why they are chosen in this analysis is given in Section \ref{subsec:compare}. In contrast to decision tree ensembles and deep learning methods, log-linear regression seeks to find a single equation to parameterize the entirety of a dataset. Deep learning has had particular success with data like images, text and audio \cite{GoodBengCour16} but has yet to see the same success with structured data in tabular form where decision trees still dominate \cite{Deep_Learning_is_not_all_you_need}.

However, tree-based learning approaches do not make use of differentiable gradients during their construction, meaning they cannot be used in component pipelines combining different models and individual modules.  Here, deep learning has the potential to create end-to-end pipelines for problems, where some of the inputs could come from tabular data and others from raw diagnostic measurements or images, and a full deep learning model could be trained in one computational graph. This, as well as the potential for improved performance, has led to many proposed deep learning solutions specific for tabular data. In this work, we compare the state-of-the-art deep learning architectures designed for tabular data with popular decision tree ensemble techniques. 

Unlike log-linear regression, other ML models can efficiently utilize categorical variables. In the case of pedestal databases, these variables can be specific experimental setups of a shot, e.g., the divertor configuration, which can take on one of 6 values written in Table \ref{tab:parameter_domains}. The deep networks analyzed in this paper make use of varying approaches to extract meaningful relationships within the categorical variables. These approaches include embedding or attention layers \cite{Embedding}. Some decision tree based methods, such as XGBoost \cite{XGboost} or CATBoost \cite{prokhorenkova2019catboost}, are constructed such that they can natively produce a meaningful response to categorical inputs.
\section{Machine Learning Experiments}

\subsection{Comparing Models}\label{subsec:compare}
We investigate which tree-based ensemble and deep learning methods have advantages over the log-linear regression used for the problem of regression of $n_\text{e,ped}$ using the JET pedestal database. We focus on the relative performance of different models. Model-agnostic deep learning methodologies such as data augmentation, learning rate warmup or decay \citep{Warmup_example}, or pretraining \citep{Pretraining_2010}, are not used, except in TabNet \cite{TabNET}, which employs these in the recommended settings. While these practices have the potential to improve performance, our goal is to evaluate the intrinsic model. 

The Urano and Frassinetti log-linear regression models serve as a performance baseline. We then compare that performance with the following decision tree models, which have been chosen for their generally good performance on a wide variety of datasets as seen in \cite{Deep_Learning_is_not_all_you_need, Katzir2021NetDNFED, Gorishniy2021RevisitingDL}: 
\begin{enumerate}
\itemsep0em 
    \item \textbf{Random Forests (RF)} \citep{randomforest} An ensemble of decision trees, where each tree is fitted in parallel using a unique subset of the dataset (a process known as bagging) and the total forest prediction is the averaged prediction between all decision trees in the ensemble. As opposed to a log-linear regression tool trying to fit all points in the dataset at once, the RF ensemble leverages the fact that each tree fits a small subset of the total dataset, which could lead to a better generalization of the whole dataset.
    \item \textbf{Extreme Randomized Trees (ERT)}  \cite{ERT_2006} An implementation of RFs that leads to more diversified trees. This is achieved by changing the algorithm of how a node is split in a decision tree, where in the RF a split is made based on an information criterion, and in ERT's this split is made randomly.  
    \item \textbf{XGBoost} \cite{XGboost} A decision tree ensemble that can build problem specific trees to aid in the ensemble prediction. Instead of fitting decision trees in parallel like RF's and ERT's, gradient boosted trees (GBDTs) conducts tree fitting sequentially: decision trees are fit iterative (one at at time) with each new tree fit by applying the gradient of the error in the previous. In XGBoost, the error application is Newton-Raphson method in function space while a generic GBDT uses gradient descent. This could enable the model to build problem specific trees, e.g., trees dedicated to high $n_\text{e,ped}$ regression. 
    \item \textbf{CATBoost} \cite{prokhorenkova2019catboost} A GBDT implementation that leads to more balanced trees than XGBoost. In each level of the decision tree, the node split combination that leads to the best prediction is selected and used for all the nodes on that level. This could enable the model to reduce overfitting in comparison to XGBoost, while still utilizing the benefits of gradient boosting. 
\end{enumerate}

To compare against the tree-based ensembles, we investigate the following deep learning architectures designed specifically for tabular data: 

\begin{enumerate}
\itemsep0em 
    \item \textbf{Multi-layer Perceptron (MLP)} The simplest ANN to establish the basic $n_\text{e,ped}$ regression capabilities of deep learning. 
    \item \textbf{Feed-Forward Nets with Category Embedding (FFCAT)} A simple feed forward network with additional layer types that are specifically designed for categorical input parameters. With this model we can leverage the MLP capabilities while adding functionality to the model in regards to using categorial variables.
    \item \textbf{TabNet} \cite{TabNET} Once fit, this model selects which features to be used in prediction on a per-instance basis (attention layers). This could be useful when predicting $n_\text{e,ped}$, as the scaling seems to be different for low and high $n_\text{e,ped}$ ranges, as observed in the struggle of the log-linear regression in fitting both.   
    \item \textbf{AutoInt} \cite{DBLP:journals/corr/abs-1810-11921} This model can learn correlations between parameters by mapping both numerical and categorical inputs into a low-dimensional space. It also utilizes attention layers like TabNet.  
    \item \textbf{NODE} \cite{NODE_2019} An ensemble of decision trees that are fitted instead using the gradient descent methods of ANN's. Like CATBoost, this model uses  symmetric splitting of each tree, but once again in a differential form. This model leverages the benefits of decision tree ensemble methods, but has the ability to be paired with other ANN models in the case of a multi-modal pipeline. 
\end{enumerate}

\subsection{Input Space}\label{subsec:input_space}

We investigate whether the inclusion of both categorical and more numerical input parameters leads to an increase in the performance of ML models in comparison to the inputs defined in traditional log-linear scalings. Additionally, we investigate whether the inclusion of global parameters, $\beta_\text{N}, Z_\text{eff}$, also aid in prediction. A fully predictive model would be able to function without these global parameters, thus it is important to determine if regressive models can function without them.  All possible input parameters and their corresponding sub domains are listed in Table \ref{tab:parameter_domains}. Using these input parameters, we create 6 unique sets of possible inputs to a model, which are listed in Table \ref{tab:input_spaces}. For input spaces U and F, we use the input parameters from the established scaling laws from Urano and Frassinetti, respectively. 

\begin{table}[h]
\caption{List of parameters included in each input parameter space from the categories in Table \ref{tab:parameter_domains}. The parameter sets are named by which variables they include: \textbf{E}: Main Engineering, \textbf{C}: Categorical, \textbf{B}: $\beta_N$, \textbf{Z}: $Z_{\text{eff}}$. \textbf{U} and \textbf{F} refer to the Urano and Frassinetti subspaces respectively.}
    \label{tab:input_spaces}
    \centering
    \begin{tabular}{c|c c c c c c c c }
        \hline
        Parameter Sets & U & F & E & EC & ECB & ECBZ  & EB & EBZ   \\
        \hline
        Main Engineering &$\dagger$ &  $\ddagger$ & * &  * &  * &  * &  * &  * \\ 
        Categorical  &  &  & & * &  * &  * &   &   \\
        $\beta_N$ & & &   &  & * & * & * & * \\
        $Z_{\text{eff}}$ & &&   &  &  & *&  & *  \\
        \hline
    \end{tabular}
    
\end{table}

\subsection{Experiment Details}
For each input space, the testing of each model is done in the following systematic approach.

\textbf{Preprocessing}. The standard scalar transformation from the Scikit-learn library \citep{scikit-learn} was applied to the numerical features as well as the target variable, $n_\text{e,ped}$. This transforms each feature to a domain with mean of zero and standard deviation of 1, which reduces the impact of the relative magnitude of each feature. The categorical features are transformed into integer representations, e.g., binary features like kicks are transformed into a 1 or 0 for a shot with and without kicks, respectively. 

The dataset is split into training-validation-test subsets, so that all model types use the same set of training data. Since the target variable, $n_\text{e,ped}$, is continuous, a pseudo stratified binning procedure is implemented such that the training-validation-test splits are even in their representation of $n_\text{e,ped}$. The resulting splits were examined using the Kolmogorov-Smirnov test from which the training, validation, and test sets all shared a p-value above 0.99, which is to say each set is likely to come from the same distribution and the null hypothesis that each set comes from the same distribution can be accepted.

The resulting sizes of the training-validation-test splits are 2116, 907, and 534 entries, respectively.

\textbf{Hyperparameter Tuning}.
It was seen that the tuning of hyperparameters played little role in improving the performance metrics of each model, therefore we abstained from hyperparameter optimization in the evaluation of models and utilized the baseline model hyperparameters suggested in each model's respective papers. Often, ML models  trained on large datasets benefit from large scale Bayesian optimization routines, but due to the relatively small size of the dataset this was left out. 

\textbf{Evaluation and Metrics}.
We run 15 experiments with different random seeds when splitting the dataset into training and validation sets, and report the Root-Mean-Squared-Error (RMSE) on the unseen test set. 

\section{Results}

\subsection{Model Comparison}
\begin{figure}[h!]

    \centering
    \includegraphics[scale=0.45]{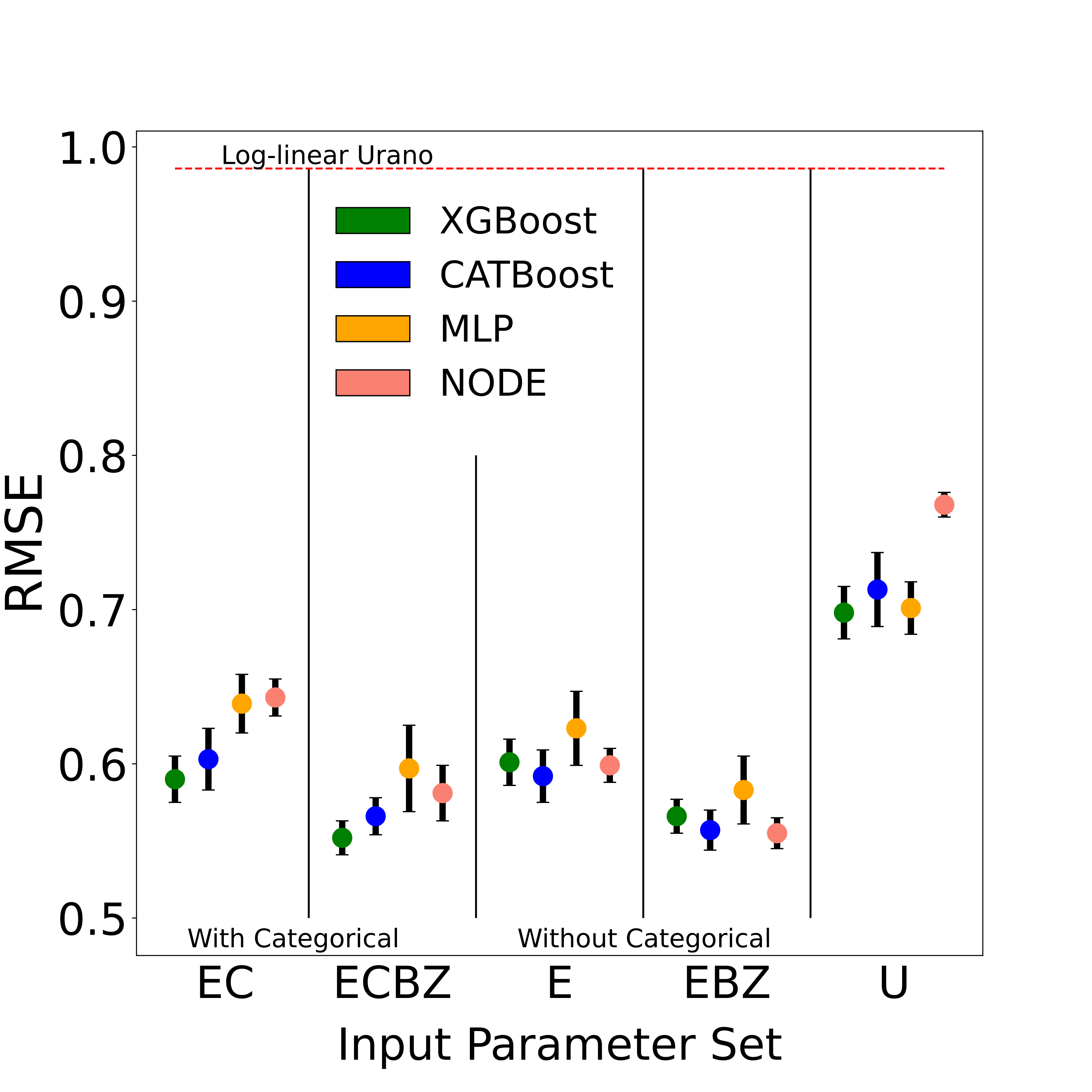}
    \caption{Results of the RMSE (on the order of $10^{19}$) of model prediction on the test set for various parameter sets and models compared to the performance of the log-linear regression used by EUROPED (dashed red line with RMSE 0.968). The average RMSE for each model is taken on the test set for 15 random splits of training/validation data. The standard deviation for the performances is visualized by the error bars. The parameter sets chosen show the differences in performance based on including/excluding categorical input parameters as well as the reduced set of input parameters used in the log-linear regression models. The models chosen are the top two performers for deep learning and decision tree ensemble groups. See Table \ref{tab:resultsstd} for a full table of results and standard deviations for all models and input spaces.}
    \label{fig:results}
    
\end{figure}

\subsubsection*{Model Performance}
\begin{figure}[h!]

    \centering
    \includegraphics[scale=0.2]{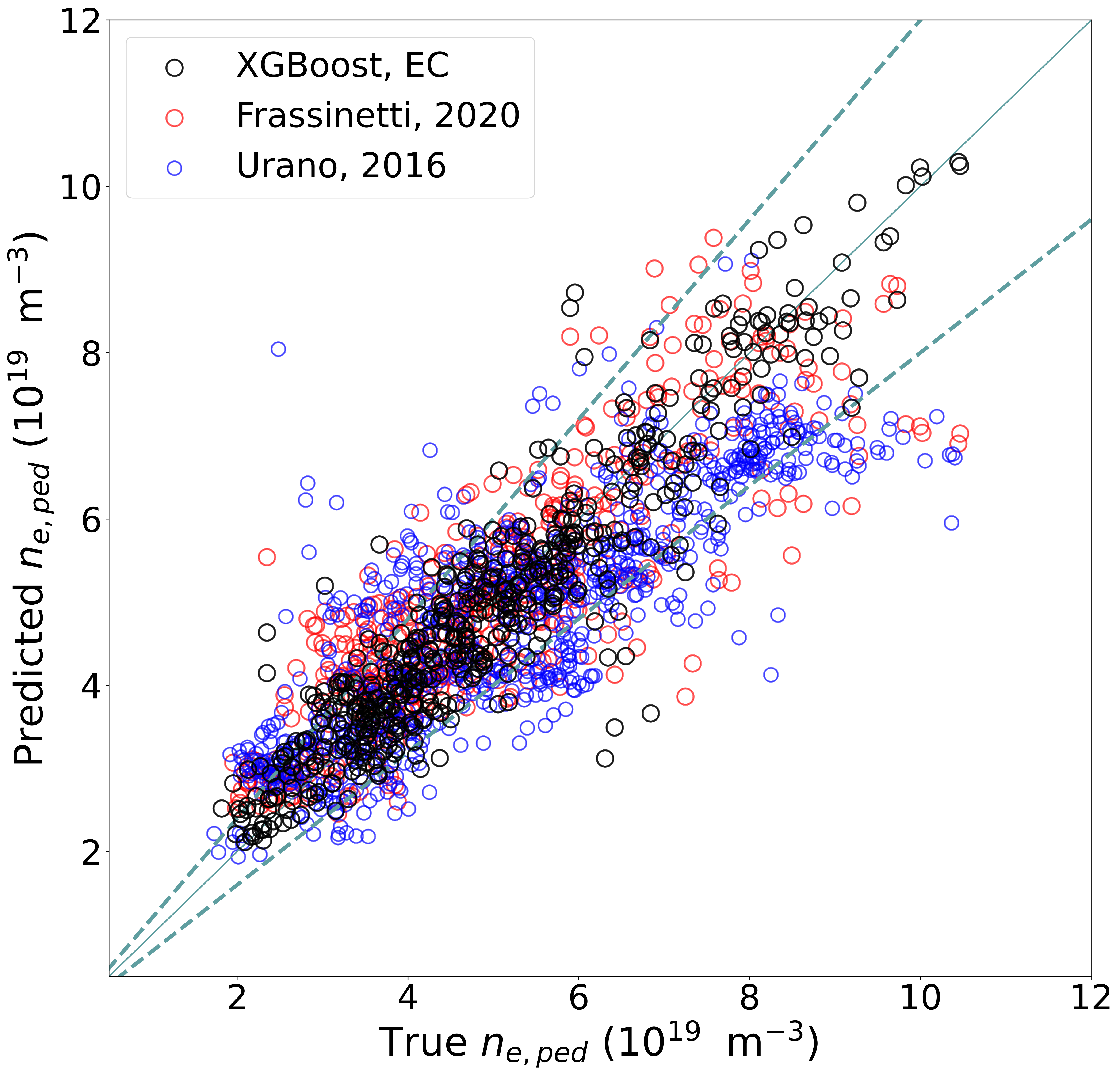}
    \caption{Predictions of $n_\text{e,ped}$ from the log-linear scaling law from Frassinetti \citep{Frassinetti_2020} and Urano \cite{Urano} (red and blue) compared to XGBoost model trained on parameter space \textbf{EC} (black). The log-linear scaling law shows systematic underprediction with experimental observations for $n_\text{e,ped} \geq 8 \times 10^{19} \; (\text{m}^{-3})$, which does not occur for XGboost. The solid line marks the perfect match and the dashed lines represent 20\% error}
    \label{fig:lin_vs_xgboost}
    
\end{figure}

Decision tree ensembles and deep learning based regression models outperform the log-linear scaling laws by about 27\% and 20\% respectively when fit using the Urano input space (Fig \ref{fig:results} Table \ref{tab:resultsstd}). This observation clearly shows the utility of these regression tools in building experimental models for physical observables that encompass highly non-linear dependencies in the plasmas. Decision tree based models perform systematically better than deep learning models with XGBoost performing the best. Of the deep learning architectures, NODE achieves the lowest RMSE and has very similar performance as the decision tree techniques when input spaces do not contain categorical features. This comes as no surprise, as NODE is essentially a neural network wrapper for GBDTs. The XGBoost shows a very good regression performance throughout the $n_\text{e,ped}$ range, including the high density points, for which the log-linear scaling laws systematically underpredict the $n_\text{e,ped}$ values (Fig. \ref{fig:lin_vs_xgboost}). 

\textbf{Takeaway:} If sufficient experimental data is available to build an interpolating regression model for $n_\text{e,ped}$, as is the case for many present-day tokamaks, the results indicate that using GBDT-type regression approaches is likely to outperform the conventional log-linear regression approach. 

\subsection{Input Spaces}
Increasing input parameters to all available engineering control parameters the prediction quality is further improved by about 13\% for both decision tree ensembles and deep learning models with respect to models fit using the Urano input space. This suggests that additional information in the form of shaping relations $a, \kappa, q_{95}, V_P$ are important to $n_\text{e,ped}$ predictions. 

When categorical variables are introduced, the change in prediction quality is negligible for decision tree ensembles and NODE, but leads to an average degradation by 3\% for other deep learning models. The lack of improvement suggests the additional categorical variables have less impact on $n_\text{e,ped}$ than the numerical variables, which is surprising as divertor configuration and impurity seeding, for example, should impact $n_\text{e,ped}$ prediction. The decrease in quality of predictions from deep learning models may come from the categorical variables competing for relevance in the model's attention layers, which are normally utilized with more leading order numerical parameters.

Including global plasma quantities $\beta_N$ and $Z_{eff}$ leads to a further 5\% increase in prediction quality for both decision tree ensembles and deep learning models. An improvement in prediction quality is expected, since $\beta_N$ plays a role in pedestal performance and encodes density information via the global plasma pressure. However, the small improvement relative to only using machine parameters as inputs suggests that most of the relevant information about $n_\text{e,ped}$ is contained in the machine parameters, and would be a small sacrifice in order to maintain a fully predictive model.

\textbf{Takeaway:} We suggest to increase the input space to all available numerical machine parameters for a given surrogate model. The inclusion of $\beta_N$ and $Z_{eff}$ lead to small improvements but renders the model not fully predictive.

\subsection{Computational Efficiency}

\begin{table}[h!]
    \caption{For selected models, the following statistics are collected: the training time (Training) and the amount of predictions per second when performing inference for each model (Throughput). These experiments use input space \textbf{EC} from Table \ref{tab:input_spaces}, which has 17 total features. Notation: $\downarrow $ ~ lower is better, $\uparrow $ ~ higher is better. Models with GPU compatibility are trained on one or more Nvidia V100s and others used 40 Intel Xeon processors. Inference timings were done using the Intel processors to eliminate differences in final performance.}
    \label{tab:timings}
    \centering
    \begin{tabular}{c|c c c c c}
        Model & Random Forest & XGBoost & MLP & TabNet & NODE  \\
        \hline
        Training (s) $\downarrow$&  $<$1 & 5 & 10 & 465 & 792 \\ 
        Throughput (pred/s) $\uparrow$& 50 & 850 & 7400 & 230 & 150 \\ 
        \hline
    \end{tabular}
    
\end{table}

We observed that the decision tree based techniques required 99\% less time to train than the more complex deep learning methods (\ref{tab:resultsstd}). All models are capable of producing more than 50 predictions per second and thus could be fast viable components of a surrogate model pipeline.

\textbf{Takeaway:} We continue to suggest GBDT methods like XGBoost or CATBoost if a simple surrogate model is desired. When an end-to-end deep learning model is suitable, e.g., model pipelines consisting of other gradient based ANNs, we suggest NODE.   
\begin{table}[h]
\resizebox{\textwidth}{!}{%
    \centering
    \begin{tabular}{c|c c c c c c c c }
        \hline
        \multicolumn{9}{c}{Input Spaces} \\
        \hline
        Model & EC & ECB & ECBZ & E & EB & EBZ & U & F \\
        \hline
        \multicolumn{9}{c}{Decision Trees} \\
        \hline
        RF & 0.642 $\pm 0.012 $& 0.634 $\pm 0.012 $& 0.624 $\pm 0.015 $& 0.66 $\pm 0.012$& 0.637 $\pm 0.012 $& 0.621 $\pm 0.013 $& 0.733 $\pm 0.017 $& 0.732 $\pm 0.019 $\\
        ERT & 0.62 $\pm 0.018$ & 0.606 $\pm 0.018$& 0.58 $\pm 0.018$& 0.601 $\pm 0.015$& 0.592 $\pm 0.015$& 0.568 $\pm 0.015$& \textbf{0.682} $\pm 0.025$& \textbf{0.674}  $\pm 0.021$\\  
        XGBoost & \textbf{0.59} $\pm 0.015$& \textbf{0.574} $\pm 0.017$& \textbf{\textcolor{teal}{0.552}} $\pm 0.011$& 0.601 $\pm 0.015$&0.589 $\pm 0.014$ & 0.566 $\pm 0.011$& 0.698 $\pm 0.017$& 0.706 $\pm 0.015$\\ 
        CATBoost & 0.603 $\pm 0.02$& 0.598 $\pm 0.018$&  0.566 $\pm 0.012$& \textbf{0.592} $\pm 0.017$& \textbf{0.584} $\pm 0.013$& 0.557 $\pm 0.013$& 0.713 $\pm 0.024$& 0.703$\pm 0.017$ \\
        \hline
        \multicolumn{9}{c}{Deep Learning} \\
        \hline 
        MLP & 0.639 $\pm 0.019$&  0.60 $\pm 0.02$&  0.597 $\pm 0.028$& 0.623 $\pm 0.024$& 0.633 $\pm 0.02$& 0.583 $\pm 0.022$& 0.701 $\pm 0.017$& 0.760$\pm 0.012$ \\  
        FFCAT & 0.672 $\pm 0.018$& 0.663 $\pm 0.021$ & 0.657 $\pm 0.026$& 0.655 $\pm 0.022$& 0.624 $\pm 0.015$& 0.623 $\pm 0.021$& 0.756 $\pm 0.014$& 0.763 $\pm 0.017$\\ 
        TABNET & 0.829 $\pm 0.041$ & 0.849 $\pm 0.038$& 0.808  $\pm 0.043$ & 0.83 $\pm 0.04$& 0.84 $\pm 0.047$&  0.775 $\pm 0.037$& 0.845 $\pm 0.022$ &  0.855 $\pm 0.05$\\ 
        AUTOINT & 0.723 $\pm 0.027$ & 0.71 $\pm 0.021$& 0.722 $\pm 0.021$ & 0.696 $\pm 0.021$ & 0.68 $\pm 0.024$& 0.69 $\pm 0.023$ &  0.816 $\pm 0.026$& 0.832 $\pm 0.032$ \\
        NODE & 0.643 $\pm 0.012$ & 0.621 $\pm 0.014$& 0.581 $\pm 0.018$& 0.599 $\pm 0.011$& 0.586 $\pm 0.012$& \textbf{0.555}$\pm 0.01$ & 0.768 $\pm 0.008$& 0.769 $\pm 0.007$\\ 
    \end{tabular}}
    \caption{Results of experiments, where \textbf{bold} numbers represent the best performance (root-mean-squared-error and standard deviation on the order of $10^{19}$) for each input space, and \textbf{\textcolor{teal}{green}} denotes the best performance overall. The log-linear scaling laws for Urano and Frassinetti obtained a RMSE of 0.986 and 0.942 respectively.}
    \label{tab:resultsstd}
\end{table}

\section*{Conclusion}

In this work, decision tree ensembles and deep learning algorithms were explored for building regression models for $n_\text{e,ped}$ at JET. Based on the model experiments conducted on the EUROfusion JET pedestal database \cite{Frassinetti_2020}, decision tree ensembles and deep learning methods are viable alternatives for surrogate modeling of $n_\text{e,ped}$ using the JET pedestal database. Both decision tree ensembles and deep learning architectures outperform log-linear regression by 23\% when applying the same set of input parameters as used in log-linear regressions by Frassinetti \cite{Frassinetti_2020} and Urano \cite{Urano}. Extending the input space to more available machine control parameters, some of which are strongly cross-correlated, we achieve an additional 13\% improvement. Including global parameters, such as $\beta_N$ and $Z_{eff}$, to the input space, improves prediction quality further by 5\% but also renders the regression model not fully predictive. Unexpectedly, the categorical parameters, such as divertor configuration or atomic number of seeded impurity, did not seem to impact the prediction quality significantly, even though divertor configuration is know to impact pedestal density in tokamaks \cite{Giroud_2015}. The detailed investigations for the reasons for this observation are a subject for future studies.

This work can serve as a basis for further developments of using decision tree based or deep learning models as alternatives for $n_\text{e,ped}$ regression in pedestal models, such as EUROPED. 


\section*{Acknowledgments}
This work has been carried out within the framework of the EUROfusion Consortium, funded by the European Union via the Euratom Research and Training Programme (Grant Agreement No 101052200 — EUROfusion). Views and opinions expressed are however those of the author(s) only and do not necessarily reflect those of the European Union or the European Commission. Neither the European Union nor the European Commission can be held responsible for them.

The authors wish to acknowledge CSC – IT Center for Science, Finland, for computational resources.

The foundations of this paper were done as part of the B.Sc. thesis of Adam Kit, which was supervised by Prof. Mathias Groth of Aalto University and Prof. Frank Cichos of Leipzig Universität, and therefore the authors extend their thanks to these two for their help in that process. 

The authors wish to thank Yoeri Poels and Andreas Gillgren for the useful conversations and contributions to this project. 

\bibliography{sample}
\newpage

\end{document}